\documentclass[amsmath,aps,superscriptaddress,prb,twocolumn,showpacs,citeautoscript]{revtex4}
\usepackage{graphicx}
\usepackage[utf8]{inputenc}
\usepackage[T1]{fontenc}
\usepackage{xspace}

\DeclareGraphicsExtensions{.pdf,.eps}

\newcommand{\wn}{\ensuremath{~\text{cm}^{-1}}\xspace}

\newcommand{\tsub}[1]{$_{#1}$}

\begin{document}

\title{Exciton-magnon transitions in the frustrated chromium antiferromagnets
CuCrO$_2$, $\alpha$-CaCr$_2$O$_4$, CdCr$_2$O$_4$, and ZnCr$_2$O$_4$}
\author{M.~Schmidt}
\author{Zhe Wang}
\author{Ch.~Kant}
\author{F.~Mayr}
\affiliation{Experimental Physics~V, Center for Electronic
Correlations and Magnetism, University of Augsburg,
D-86135~Augsburg, Germany}

\author{S.~Toth}
\affiliation{Helmholtz Zentrum Berlin f{\"u}r Materialien und
Energie, D-14109~Berlin, Germany}

\author{A.T.M.N.~Islam}
\affiliation{Helmholtz Zentrum Berlin f{\"u}r Materialien und Energie, D-14109~Berlin, Germany}

\author{B.~Lake}
\affiliation{Helmholtz Zentrum Berlin f{\"u}r Materialien und Energie, D-14109~Berlin, Germany}
\affiliation{Institut f{\"u}r Festk{\"o}rperphysik, Technische Universit{\"a}t Berlin, D-10623~Berlin, Germany}

\author{V.~Tsurkan}
\affiliation{Experimental Physics~V, Center for Electronic
Correlations and Magnetism, University of Augsburg,
D-86135~Augsburg, Germany} \affiliation{Institute of Applied
Physics, Academy of Sciences of Moldova, MD-2028~Chisinau, Republic
of Moldova}

\author{A.~Loidl}
\affiliation{Experimental Physics~V, Center for Electronic
Correlations and Magnetism, University of Augsburg,
D-86135~Augsburg, Germany}

\author{J.~Deisenhofer}
\affiliation{Experimental Physics~V, Center for Electronic
Correlations and Magnetism, University of Augsburg,
D-86135~Augsburg, Germany}

\date{\today}

\begin{abstract}
We report on optical transmission spectroscopy of the Cr-based
frustrated triangular antiferromagnets CuCrO$_2$ and
$\alpha$-CaCr$_2$O$_4$, and the spinels CdCr$_2$O$_4$ and
ZnCr$_2$O$_4$ in the near-infrared to visible-light frequency range. We explore the possibility to search for spin correlations far above the
magnetic ordering temperature and for anomalies in the magnon lifetime in the magnetically ordered state by probing exciton-magnon sidebands of the spin-forbidden crystal-field transitions of the Cr$^{3+}$ ions (spin $S$ = 3/2). In CuCrO$_2$ and $\alpha$-CaCr$_2$O$_4$ the appearance of fine structures below $T_\text{N}$ is assigned to magnon sidebands by comparison with neutron scattering results. The temperature dependence of the line width of the most intense sidebands in both compounds can be described by an Arrhenius law. For CuCrO$_2$ the sideband associated with the $^4$\textit{A}\tsub{\mathsf{2}}
$\rightarrow ^2$\textit{T}\tsub{\mathsf{2}} transition can be observed even above $T_\text{N}$. Its line width does not show a kink at the magnetic ordering temperature and can alternatively be described by a $Z_2$ vortex scenario proposed previously for similar materials. The exciton-magnon features  in $\alpha$-CaCr$_2$O$_4$ are more complex due to the orthorhombic distortion. While for CdCr$_2$O$_4$ magnon sidebands are identified below $T_\text{N}$ and one sideband excitation is found to persist across the magnetic ordering transition, only a weak fine structure related to magnetic ordering has been observed in ZnCr$_2$O$_4$ .

\end{abstract}


\maketitle

\section{Introduction}

An archetype of a geometrically frustrated spin arrangement are
antiferromagnetically coupled Ising spins residing on the corners of
a triangular lattice. In three dimensions the pyrochlore lattice
which can be regarded as a network of corner-sharing tetrahedra is
one of the most studied structures to explore the realm of
frustration phenomena in magnetism.

These geometrically frustrated lattices may
provide evidence for exotic ground states such as spin ice, spin liquids,
or spin gels. The latter has been described as a topologically ordered state with
finite but extended spin correlations and was proposed to be realized
in systems with antiferromagnetically coupled Heisenberg spins on a
two-dimensional (2D) triangular lattice.\cite{Kawamura2010} In this
case the nearest-neighbor bilinear interaction yields a long-range ordered magnetic state at $T=0$~K, where neighboring spins order at an angle of 120$^\circ$ to each other.\cite{Bernu1994}
Moreover, the model exhibits a topologically stable defect described in terms of a
\textit{Z}\tsub2 vortex \cite{Kawamura1984}.

A well-known class of triangular antiferromagnets is given by
systems with chemical formula $A$CrO$_2$ with $A$ = Cu, Ag, Pd, Li,
Na. Depending on the stacking sequence these compounds have either
delafossite structure (e.g.\ Cu/PdCrO$_2$) or an ordered rock salt
type structure (e.g.\ Li/NaCrO$_2$). These systems have attracted
enormous interest because of a large variety of magnetic and
electronic phenomena and the occurrence of multiferroicity. For
example, CuCrO$_2$ and AgCrO$_2$ reportedly exhibit spin-driven
ferroelectricity \cite{Seki2008} and the spin correlations in both
rock salt- and delafossite-type $A$CrO$_2$ have been associated with
the formation of \textit{Z}\tsub2 vortices \cite{Ajiro1988,
Olariu2006, Hsieh2008, Hsieh2008a, Hemmida2009,Hemmida2011}.


\begin{figure*}[t]
\includegraphics[width=0.8\textwidth]{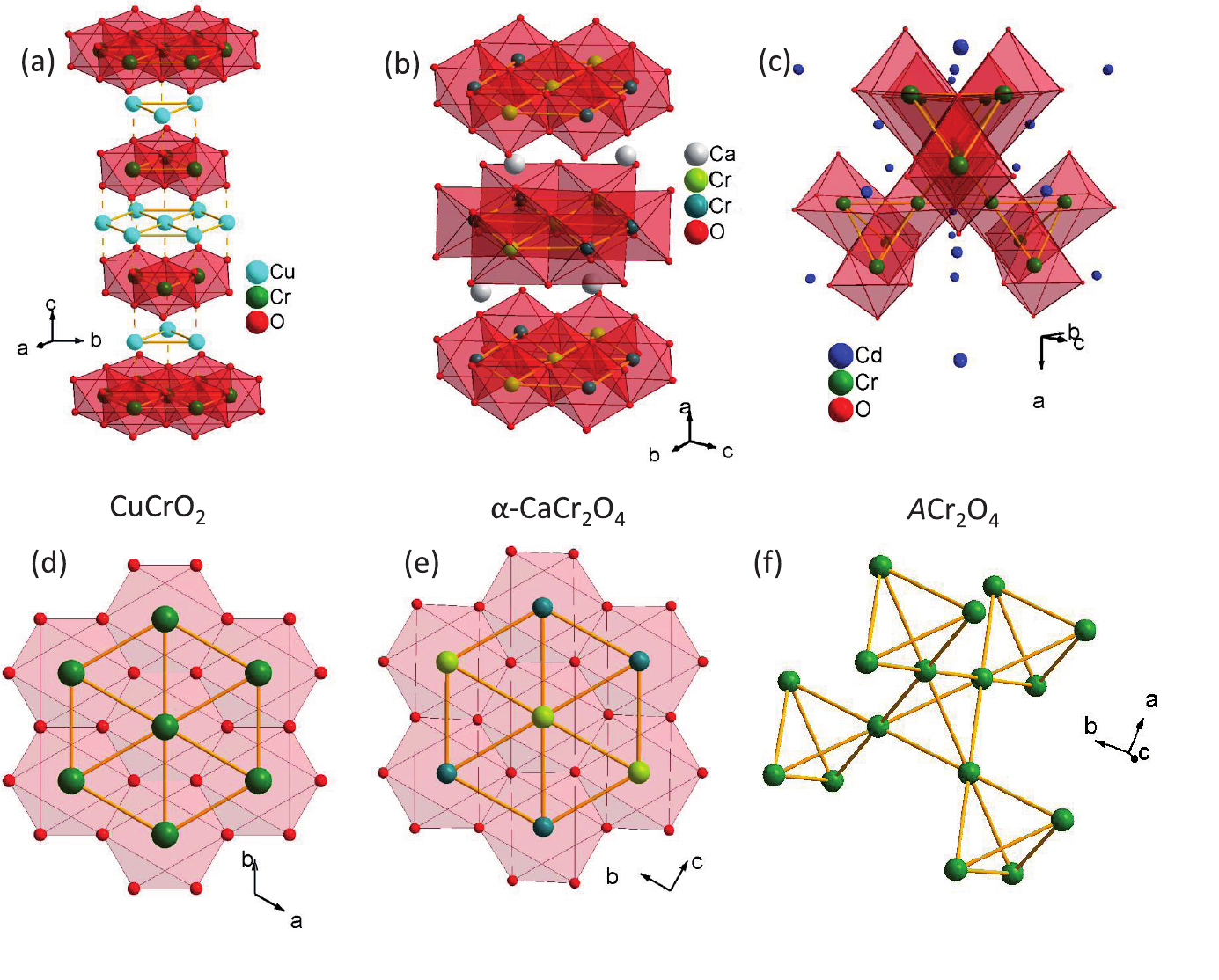}
\caption{\label{fig:struc} (Color online) Room temperature crystal structures of (a)
CuCrO$_2$, (b) $\alpha$-CaCr$_2$O$_4$, and (c) $A$Cr$_2$O$_4$ with
$A$=Cd, Zn. In (d)-(f) we show the respective triangular planes and
the pyrochlore lattice formed by the Cr ions in these compounds.}
\end{figure*}

Optical studies have been reported in the rocksalt-type systems
NaCrO$_2$ and LiCrO$_2$ \cite{Elliston1975,Kojima1993}. These
systems exhibit exciton-magnon transitions\cite{Tanabe1965} in the
energy region of the spin-forbidden $^4$\textit{A}\tsub{\mathsf{2}}
$\rightarrow ^2$\textit{T}\tsub{\mathsf{2}} crystal-field
excitation of the Cr$^{3+}$ ions. Kojima \textit{et al}.\ proposed
that the lifetime of the exciton-magnon lines in triangular lattice
antiferromagnets is directly related to the density of
\textit{Z}\tsub2 vortices in the system.\cite{Kojima1993} In this
work we investigated the delafossite-type system CuCrO$_2$ by
optical spectroscopy. CuCrO$_2$ crystallizes in space group
$R\overline{3}m$\cite{Kadowaki1990}  where layers of the magnetic
chromium ions (3$d^3$, $S=3/2$) are separated by one
copper and two oxygen  layers [see Fig.~\ref{fig:struc}(a) and (d)]. Due to this
distance the system is moderately frustrated with a
Curie-Weiss-temperature $\Theta_\text{CW}$ of $-190$~K  and an antiferromagnetic
ordering below $T_\text{N2}$ = 24.2~K. The magnetic transition leads
to a slight distortion of the CrO$_6$ octahedra\cite{Poienar2009,
Kimura2009} and an incommensurate proper screw magnetic
structure.\cite{Soda2009} Specific heat measurements showed that
there are actually two successive phase transitions at
$T_\text{N1}$= 23.6~K and $T_\text{N2}$= 24.2~K.\cite{Kimura2008} At
$T_\text{N2}$ a two-dimensional antiferromagnetic ordering was
proposed and only below $T_\text{N1}$ three-dimensional magnetic
ordering and the occurrence of multiferroicity sets
in.\cite{Frontzek2011} In the following we will use $T_\text{N}=T_\text{N1}=23.6$~K because our measurements do not allow to distinguish these two transitions.

Moreover, we investigated the triangular-lattice antiferromagnet
$\alpha$-CaCr$_2$O$_4$, which has an orthorhombically distorted
delafossite structure (space group $Pmmn$) \cite{Pausch1974}, where the Cr$^{3+}$ ions occupy two crystallographically inequivalent positions [see Fig.~\ref{fig:struc}(b) and (e)]. With a Curie-Weiss temperature of $-564$~K and a N\'eel temperature of $T_\text{N}$ = 42.6~K the system is clearly geometrically frustrated and exhibits a planar 120$^\circ$-spin structure in the crystallographic $ac$-plane.\cite{Chapon2011,Toth2011} Recent studies report multiferroicity (also in $\alpha$-$M$Cr$_2$O$_4$, $M$ = Sr, Ba)\cite{Singh2011,Zhao2012} and low-lying magnetic modes with a roton-like dispersion in $\alpha$-CaCr$_2$O$_4$.\cite{Toth2012} The related compound $\alpha$-SrCr$_2$O$_4$ reportedly exhibits similar magnetic properties but is less distorted than $\alpha$-CaCr$_2$O$_4$.\cite{Dutton2011}

The spinel systems CdCr$_{2}$O$_{4}$ and ZnCr$_{2}$O$_{4}$ with a pyrochlore lattice of magnetic Cr$^{3+}$ ions [see
Fig.~\ref{fig:struc}(c) and (f)] are considered as model systems to study the
effects of geometric frustration of Heisenberg spins on the
pyrochlore lattice. Further-neighbor exchange interactions \cite{chern08} and
magneto-elastic coupling lead to magneto-structural transitions
\cite{yamashit00,tchernys02,sushkov05,fennie06,aguilar08,rudolf09a,kant09a,Kant2012}at N\'eel temperatures of 12.5~K and 7.8~K  for ZnCr$_{2}$O$_{4}$ and CdCr$_{2}$O$_{4}$, respectively
\cite{rovers02,lee00}, while the respective Curie-Weiss
temperatures are $-390$~K and $-71$~K.\cite{rudolf07,sushkov05} The observation of magnetic excitations by
neutron scattering which can be modelled by structure factors
corresponding to a partition of the pyrochlore lattice into
hexagonal loops or even heptamers\cite{lee02,chung05,tomiyasu04}
has contributed considerably to this paradigmatic status.
A low-temperature optical absorption spectrum of ZnCr$_{2}$O$_{4}$ has been reported by Szymczak \emph{et al}\cite{Szymczak80}., where an exotic multiplet-sideband assigned to exciton-magnon-phonon processes has been observed. A clear suppression of this sideband has recently been reported in ultra-high magnetic fields up to 600~T.\cite{Miyata2011}

In this study we will focus on the exciton-magnon
transitions related to spin-forbidden CF excitations of the
Cr$^{3+}$ ions with spin $S$ = 3/2, which are in an octahedral
environment in all considered compounds.

\begin{figure*}[htbp]
  \begin{minipage}{0.75\textwidth}
    \includegraphics[width=\textwidth]{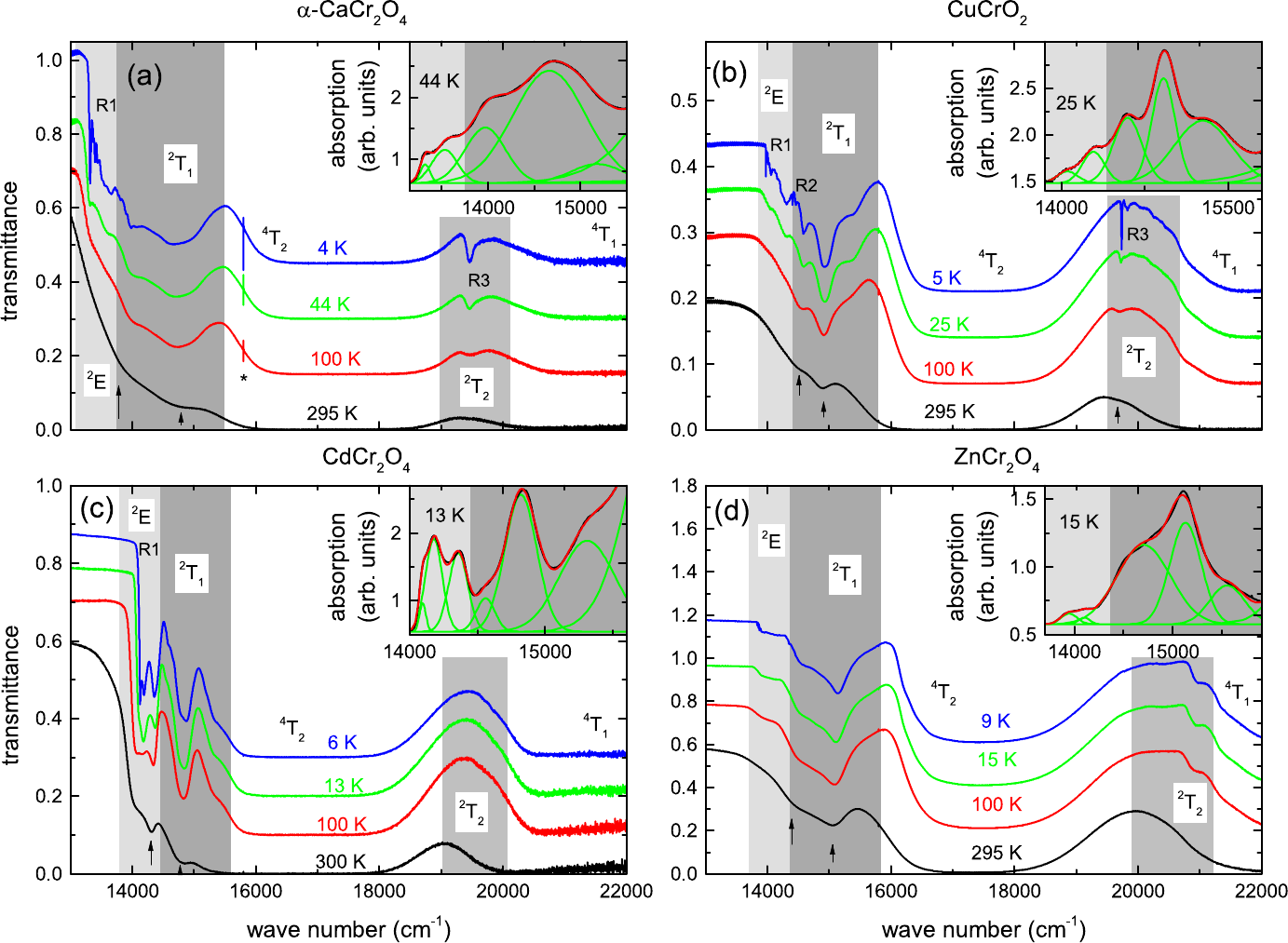}
  \end{minipage}
  \begin{minipage}{0.24\textwidth}
    \includegraphics[width=\textwidth]{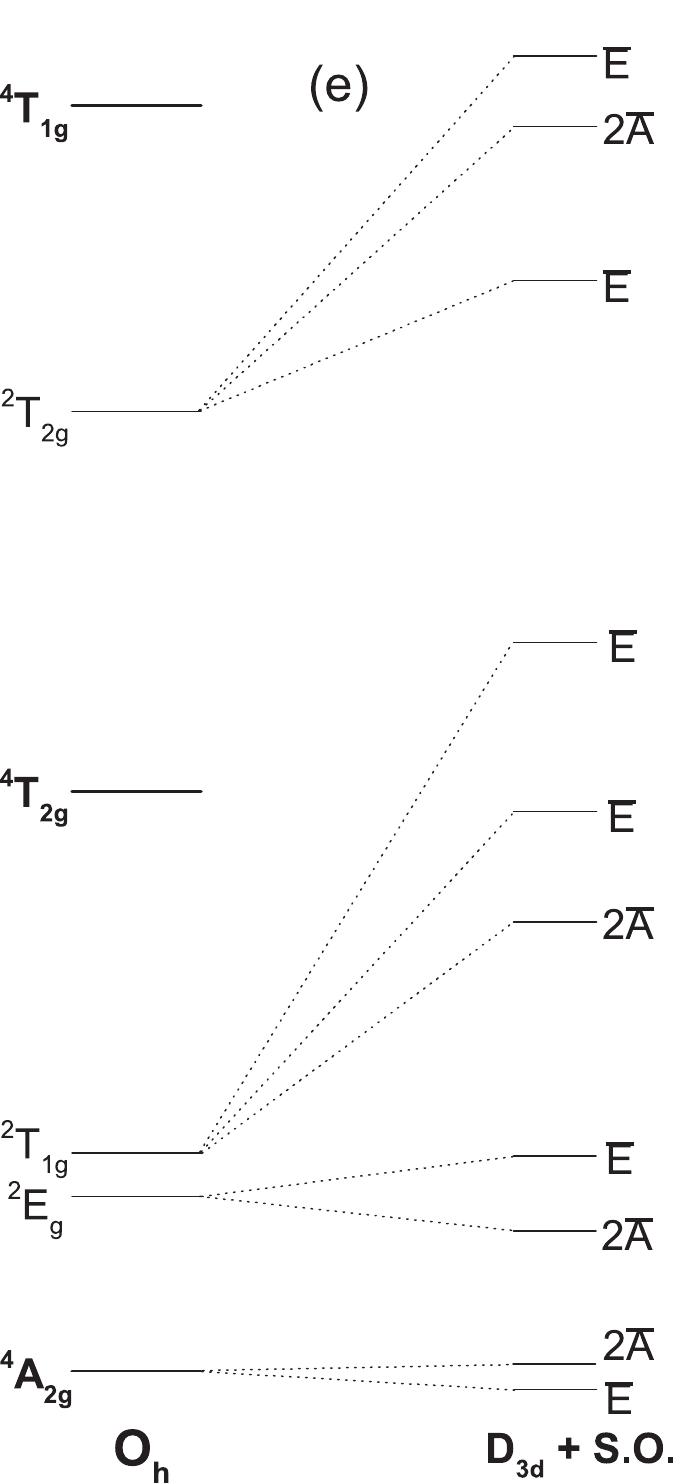}
  \end{minipage}
 \caption{\label{fig:transmission} (Color online) Transmission spectra for
 $\alpha$-CaCr$_2$O$_4$ (a), CuCrO$_2$ (b),  and  $A$Cr$_2$O$_4$ with
$A$=Cd (c), Zn (d) for several temperatures below and above the respective
magnetic ordering temperatures. The lines are shifted for clarity with respect to the highest temperature.
Up arrows mark the spin-forbidden transitions (shaded areas) already visible at room
temperature and the asterisk in (a) an artifact due to the laser of the
spectrometer. The insets in the top right corner of the frames are zooms into the lower energy absorption
spectrum just above the magnetic ordering together with a fit to the data. Individual peaks
are shown in green, while the sum is in red color. Frame (e) shows the transition scheme of a $d^3$-ion in an octahedral crystal field O$_h$ and the splitting of the spin-forbidden modes due to the trigonal environment D$_{3d}$ and spin-orbit coupling (S.O.) following Ref.~[\onlinecite{Wood68}].}
\end{figure*}

\section{Experimental details and sample characterizations}


Single crystals of CuCrO$_2$ were prepared by a flux-decomposition method using
K$_2$Cr$_2$O$_7$ flux and CuO. The soaking temperature was
1150$^\circ$C, the soaking time 20 h and the cooling rate
2$^\circ$C/h. Plate-like single-crystalline samples with dimension
up to $5 \times 5 \times 0.7$~mm$^3$ were prepared. The grown 
single crystals were checked by x-ray diffraction and no impurity
phases could be detected. The magnetic susceptibility of our samples
(not shown) is in good agreement with literature
\cite{Kimura2008, Okuda2005}. A Curie-Weiss fit to the data between
200 and 400~K results in a Curie-Weiss temperature of
178~K, 
slightly lower than the
values measured by Kimura \emph{et al}\cite{Kimura2008}.\ (211~K out
of plane, 203~K in plane) from samples grown from Bi$_2$O$_3$ flux.
The propagation direction of the incident light was parallel to the crystallographic $c$-axis.

Single crystals of $\alpha$-CaCr$_2$O$_4$ were grown in a high temperature floating zone furnace as described elsewhere \cite{Islam2012}. The cleaved platelet is about 100~microns thick with a diameter of 3~mm. The propagation direction of the incident light was parallel to the crystallographic $a$-axis.

High-quality platelike single crystals of CdCr\tsub2O\tsub4 and
ZnCr\tsub2O\tsub4 were prepared as described in
Ref.~[\onlinecite{Kant09}] and polished to optical quality. The propagation direction of the incident light was parallel to the [111]-direction.

The optical transmission was measured using a Bruker IFS 66v/S
Fourier-transform spectrometer, which was equipped with a He-bath
and a He-flow cryostat, in the frequency range
8500 - 25000~cm$^{-1}$ and for temperatures from 5 - 500 K.



\section{Experimental Results and Discussion}

\subsection{Crystal-field splitting of Cr$^{3+}$}

\begin{table}
\begin{center}
	\begin{tabular}{ l | c | c |c |c }
  		 	& ~$\alpha$-CaCr$_2$O$_4$~ 	& ~CuCrO$_2$~ 	& ~CdCr$_2$O$_4$~ 		& ~ZnCr$_2$O$_4$~	\\  \hline
	Mode 1	& 13317.7				& 14053.0		& 14096.0			& 13946.7		\\
	Mode 1*	&					&			& 14178.3			& 			\\
	Mode 2	& 13530.6				& 14286.4		& 14358.1			& 14099.5		\\
	Mode 3	& 13969.0				& 14593.3		& 14559.3			& 14702.1		\\
	Mode 4	& 14663.4				& 14916.5		& 14819.8			& 15126.7		\\
	Mode 5	& 15179.4				& 15268.9		& 15309.7			& 15549.2		\\

	\end{tabular}
\caption{\label{tab:energies} Observed transitions between the $t_{2g}$ levels for the different compounds given in wave numbers (cm$^{-1}$) obtained from a fit to the data just above the magnetic ordering temperature (see insets of Fig.~\ref{fig:transmission}).}
\end{center}
\end{table}

 In all systems investigated here the magnetic ions  are Cr$^{3+}$ ions with 3$d^3$ electronic configuration (spin $S=3/2$) surrounded by an octahedron of oxygen ions. In a perfect O$_h$ symmetry, this would result in the usual splitting according to the Tanabe-Sugano diagram for $d^3$ ions\cite{Sugano1970} [see scheme in Fig.~\ref{fig:transmission}(e)]. However, the symmetry for our compounds is rather trigonal D$_{3d}$, which leads to a splitting of the excitations. Due to spin-orbit coupling we get a further splitting. Wood \emph{et al}\cite{Wood68}.\ investigated the crystal field excitations in ZnAl$_2$O$_4$ and MgAl$_2$O$_4$ doped with Cr ions both theoretically and experimentally. The Cr ions in these spinels are in a similar environment as in our case and the $^2E$ mode splits by a few wave numbers while the $^2T$ modes are split into three levels separated by a few 100\wn. For ZnGa$_2$O$_4$ doped with Cr the splitting of the $^2E$ multiplet increases to 40\wn\cite{kahan1971}, showing the sensitivity to the local symmetry. The scheme of excitations including the low symmetry and spin-orbit splitting of the spin-forbidden modes is shown in Fig.~\ref{fig:transmission}(e).

The parity selection rule which forbids transitions between the
multiplet states in cubic symmetry can be released by a static
low-symmetry crystal field with odd-parity or a corresponding odd
lattice vibration inducing a low-symmetry field. As a result
even-parity states will be mixed with odd-parity contributions and
nonvanishing matrix elements of the electric dipole moment will
occur. No such matrix elements for transitions to multiplet
states with different spin multiplicities (spin-forbidden crystal
field transitions) exist, but this spin-selection rule ($\Delta
S=0, 1$) can be released by spin-orbit coupling.\cite{Sugano1970}

The NIR transmission spectra for all materials
are shown in Fig.~\ref{fig:transmission}. The overall
features are typical for Cr$^{3+}$ ions in an octahedral crystal field
like, for example, in ruby\cite{Wood1963}. One can see two strong
spin-allowed crystal-field absorptions in all compounds at about 17000~cm$^{-1}$
and about 22000~cm$^{-1}$ corresponding to $^4A_2 \rightarrow
{}^4T_2$ and $^4A_2 \rightarrow {}^4T_1$ transitions, respectively, where one
electron is excited from the $t_{2g}$ to the $e_g$ levels. These energies correspond to a cubic crystal field splitting 10$Dq\approx$~17000~cm$^{-1}$ and a Racah parameter $B\approx$~465~cm$^{-1}$. A splitting of the spin-allowed transitions due a crystal field
lower than cubic or spin-orbit coupling can not be
resolved.
%


The focus here is on the spin-forbidden transitions which are much weaker than
the spin-allowed transitions. In cubic symmetry the spin-forbidden
excitations $^4A_2 \rightarrow {}^2E$, $^4A_2 \rightarrow {}^2T_1$
and $^4A_2 \rightarrow {}^2T_2$ are expected and already at room
temperature weaker modes (indicated by arrows in
Fig.~\ref{fig:transmission}) are visible. Towards lower temperature these
absorption features become narrower and more pronounced and just above the magnetic ordering
five distinct absorptions are identified in the region of the $^4A_2
\rightarrow {}^2E$, $^4A_2 \rightarrow {}^2T_1$ (see insets of
Fig.~\ref{fig:transmission}). This number is in accordance with the five modes expected in this region for D$_{3d}$ symmetry [Fig.~\ref{fig:transmission}(e)]. The energies of the excitations are obtained by a fit and can be read off in Table~\ref{tab:energies}.
Even for the lower symmetry of $\alpha$-CaCr$_2$O$_4$ the fit works well with 5 Gaussians in this region. In the case of CdCr$_2$O$_4$ an additional Gaussian was used to account for the shoulder at the low-energy flank of the $^2E$ absorption region resulting in modes 1 and 1$^*$.



\begin{figure*}[ht]
\includegraphics[width=0.9\textwidth,clip]{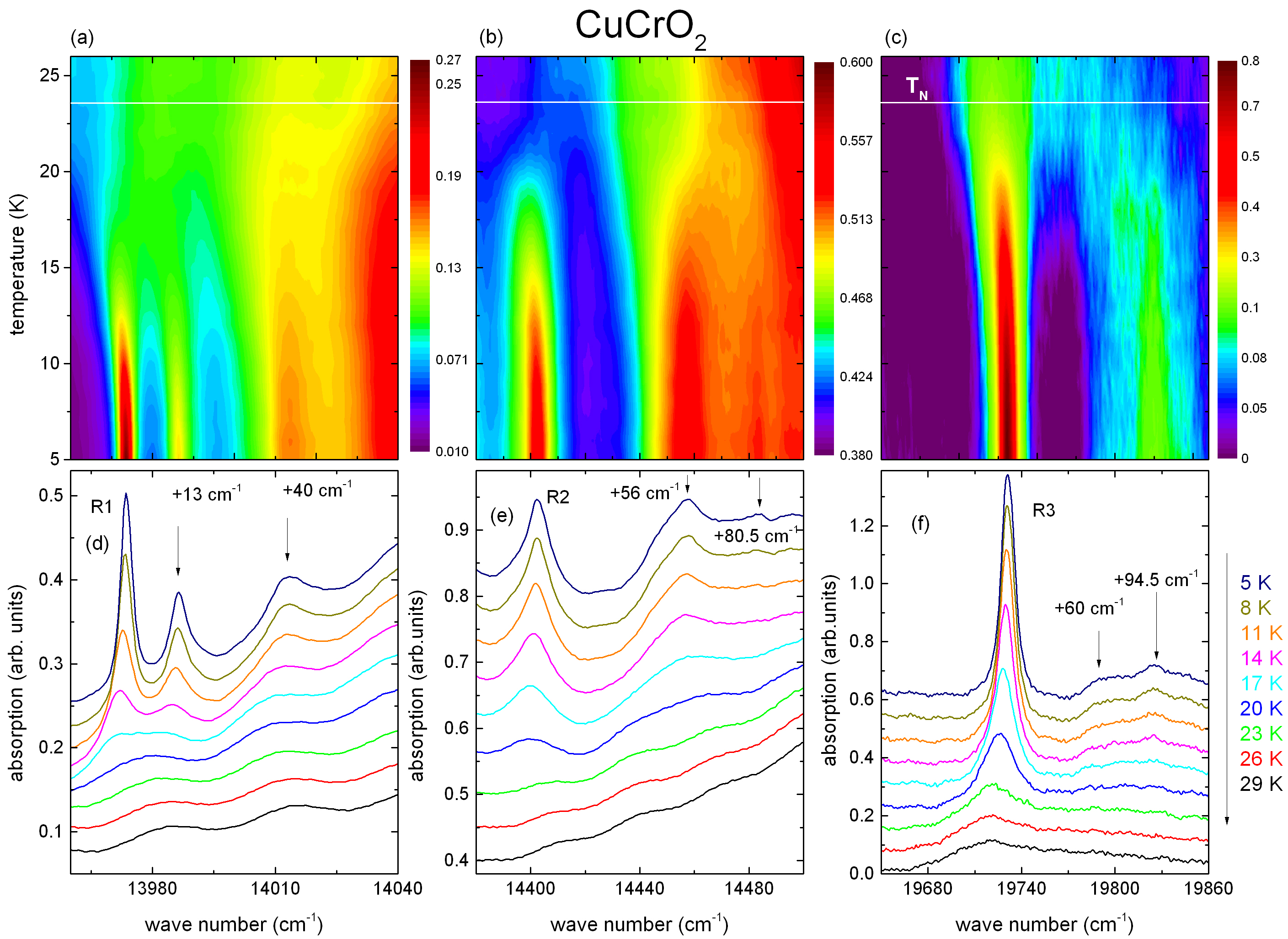}
\caption{\label{fig:range1} (Color online) Color coded contour plot
of the absorption as a function of temperature vs.\ wave number in
the range of excitation R1 (a), R2 (b), and R3 (c). The transition temperature is marked as a white line. The lower part
of the figure [(d)-(f)] shows absorption spectra in the same range
as the respective frame on top for various temperatures. The curves
are shifted for clarity by a constant value of  0.03 (d), 0.05 (e),
and 0.075 (f). Sidebands are marked by arrows together with the energy difference to R1, R2, and R3, respectively.}
\end{figure*}

\subsection{Exciton-magnon transitions}



Let us now take a closer look at the spin-forbidden crystal field
transitions within the $t_2^3$ states, in particular the
$^4A_{2}\rightarrow {}^2E_{2}$ and the $^4A_{2}\rightarrow {}^2T_{1}$
transitions. In the transmission spectra very sharp excitonic absorption features accompanying these transitions are already visible in Fig.~\ref{fig:transmission} [e.g.\ the ones labelled R1, R2, and R3 in Fig.~\ref{fig:transmission}(b)] at lowest temperatures. As proposed by Tanabe, Moriya, and Sugano such sidebands can occur when two ions on sites $i$ and $j$ are coupled
antiferromagnetically and the electric-dipole moment of the incoming light couples an excited multiplet state of the ion at site $i$ to the ground multiplet state of the ion at $j$.\cite{Tanabe1965} In this so-called electric-dipole induced exciton-magnon
process with a total spin change $\Delta S=0$, the spin-forbidden multiplet transitions can exhibit cold and hot magnon sidebands on the high- and low-energy side of the so-called zero-magnon line, corresponding to the annihihilation and creation of a magnon, respectively\cite{Gondaira1966, Tanabe67, Shinagawa1971, Fujiwara1972, Eremenko1986}. The zero-magnon line identifies the purely excitonic crystal-field transition. Electric-dipole active magnon sidebands involving spin-forbidden crystal-field levels have been reported for several antiferromagnets such as, for example, MnF$_2$,\cite{greene65, Sell1967} VBr$_2$,\cite{Kojima1993} and LiCrO$_2$,\cite{Kojima1993} and are usually stronger than the magnetic-dipole active zero-magnon lines.

In case of spin-allowed crystal-field transitions, magnetic-dipole active magnon sidebands have been observed (e.g.\ in FeF$_2$ and KCuF$_3$),\cite{tylicki1968,Deisenho08} which can be of similar strength as the electric-dipole active sidebands of spin-forbidden crystal-field transition and are not restricted to antiferromagnets.\cite{moriya1968,fujiwara1975} In contrast to the electric-dipole mechanism of spin-forbidden transitions, the zero-magnon line is expected to be stronger than the sidebands for the spin-allowed crystal-field transitions.\cite{moriya1968}

The energy of exciton-magnon sidebands with respect to the zero-magnon line should correspond to the magnon energy in regions of the Brillouin zone, where the dispersion is flat, i.e., where the magnon density of states peaks. The effective electric-dipole moment of an exciton-magnon transition and the magnon sidebands appearing for a particular excitonic transition depends strongly on the symmetry of the excited orbital state and the dispersion of the exciton.\cite{sell1968,moriya1968} As a result different magnons might couple to the different excited multiplet states. Thus, optical spectroscopy can provide information on the magnon energies in regions of the Brillouin zone which might be hard to access otherwise. Since short-range correlations are sufficient to make these transitions allowed, the persistence of non-trivial spin correlations in the classical spin-liquid phase of frustrated magnets may be tracked by optical spectroscopy. In the following we will discuss the observed exciton-magnon transitions separately for each of the investigated compounds.

\begin{figure}
\includegraphics[width=0.4\textwidth]{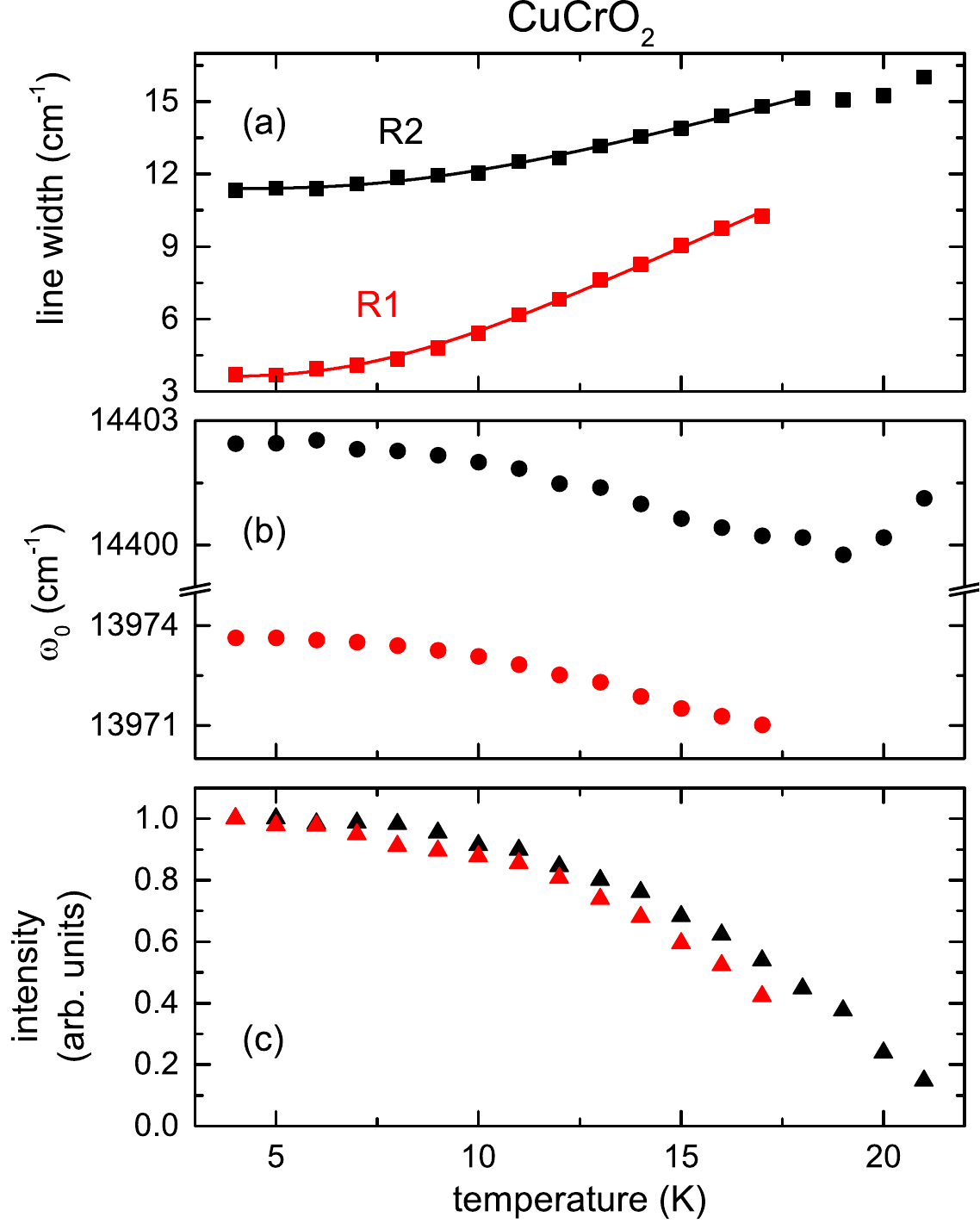}
\caption{\label{fig:fitparam_r1_r2} (Color online)  Fit parameters of the absorptions R1 (red) and R2 (black): (a) line width, (b) frequency and (c) intensity normalized to the value at 4~K. The solid lines are fits using Eq.~(1) as described in the text.}
\end{figure}

\subsubsection{CuCrO$_2$}

In the magnetically ordered state pronounced fine structures, named
R1, R2, and R3 below, appear at the onset of the
$^4A_2 \rightarrow {}^2E$, $^4A_2 \rightarrow {}^2T_1$, and  $^4A_2 \rightarrow {}^2T_2$ excitations below the magnetic
ordering temperature $T_\text{N}\approx T_\text{N1}$\cite{Kimura2008} [see Fig.~\ref{fig:range1}(a)-(f)]. In the following we will restrict the
discussion on these fine structures and their temperature dependence
which is summarized in Fig.~\ref{fig:range1}.

\begin{figure*}[t]
\includegraphics[width=0.8\textwidth]{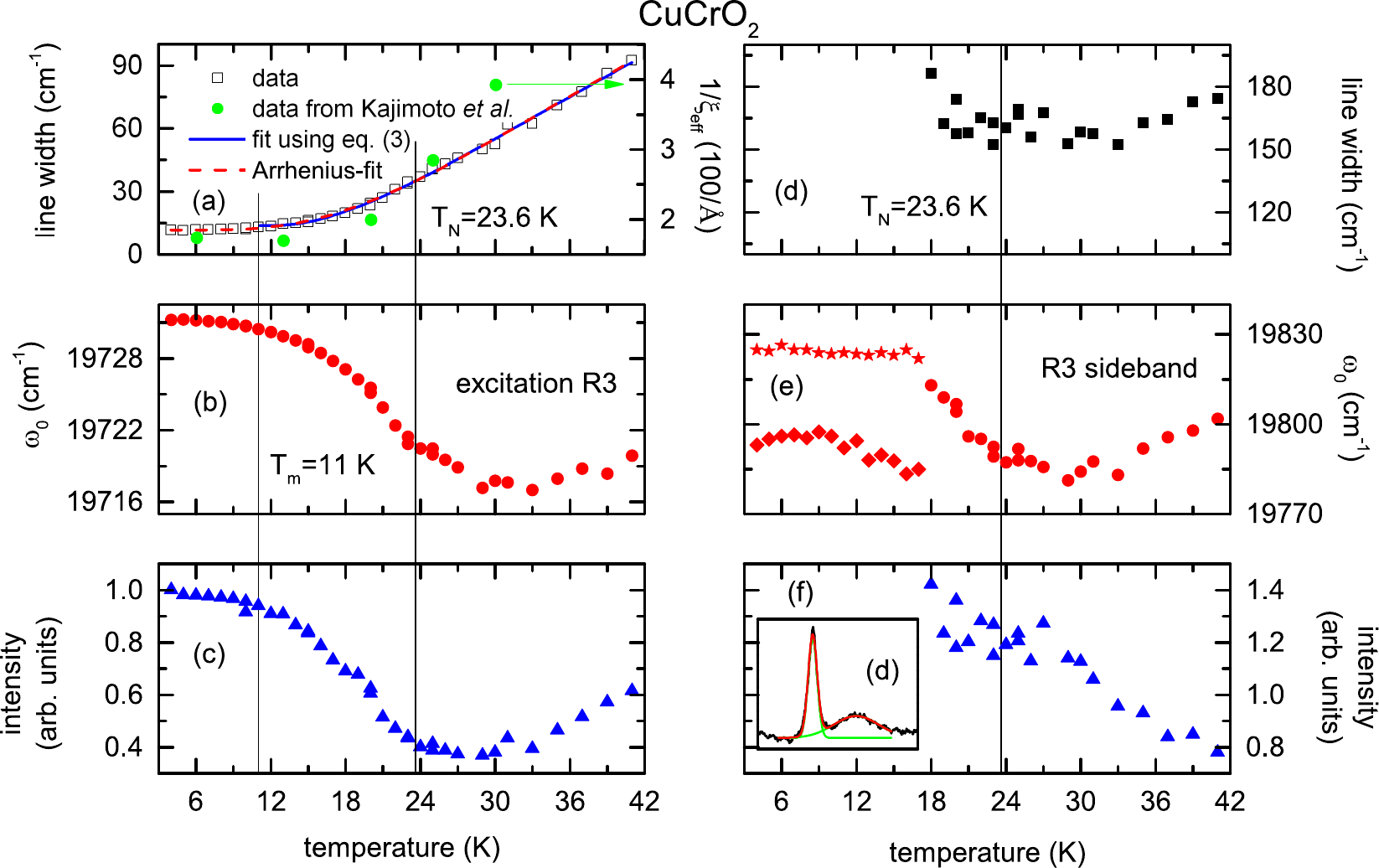}
\caption{\label{fig:fitparcucro2} (Color online)  Fit parameters of the R3 absorption (a) line width, (b) frequency and (c) intensity normalized to the value at 4~K and of the sideband of R3 (d)-(f), respectively. The red dashed and blue solid lines in (a) are fits to the line width data of excitation R3 with an Arrhenius model and a model assuming $Z_2$ vortices, respectively (see text). The green circles are data for the inverse of the effective correlation length $\xi_\text{eff}$ from Ref.~[\onlinecite{Kajimoto2010}]. Below 18~K the energies of the sideband (stars) and its shoulder (diamonds) are directly read off from the absorption data. In panel (g) an absorption spectrum at $T$=18~K is shown together with a fit using two Gaussians. The two individual peaks are in green and the sum of both in red color.}
\end{figure*}

Zooming into the region of the fine structure close to the
excitation R1 [Fig.~\ref{fig:range1}(a) and (d)] three absorption
features can be clearly seen at low temperatures. Given the spin-forbidden nature of the underlying crystal-field transition (see discussion above) the most intense peak R1 should correspond to an electric-dipole active one-magnon sideband, but no corresponding weak zero-magnon feature can be observed to directly read off the magnon energy. Consequently, the identification of the two further sidebands with an energy difference of +13\wn and +40~cm$^{-1}$ with respect to R1 is similarly difficult.

Recent high-field electron-spin resonance studies reported two antiferromagnetic-resonance (AFMR) modes with gaps of $m_1$ = 1.0~cm$^{-1}$ (0.12~meV) and $m_2$ = 11~cm$^{-1}$ (1.4~meV)\cite{Yamaguchi2010a}.  While the smaller gap has been observed also in neutron scattering studies \cite{Poienar2010,Kajimoto2010,Frontzek2011}, the larger gap has no correspondence in the investigated range of momentum transfer and energy probed in the neutron experiments. However, the energy difference of +13\wn of the magnon sideband from R1 is close in energy to the reported gap and the difference of 2\wn could be due to the fact that ESR and exciton-magnon-transition probe spin-wave energies at different points in the Brillouin zone. The small gap of only +1\wn observed by the ESR experiment and inelastic neutron scattering results in an assignment of the most intense absorption R1 to an exciton-one-magnon sideband at a distance of 1\wn to the zero-magnon line. That means the magnetic-dipole active line is assumed to be masked by R1. Then the absorption at +13\wn (1.6~meV) corresponds to an exciton-one-magnon sideband involving the larger reported gap and the sideband energy at about +40~cm$^{-1}$ (5~meV) is in agreement with reported magnon bands around $m_3$ = 5~meV
with a bandwidth of about 1~meV in the vicinity of the zone
boundary.\cite{Poienar2010,Frontzek2011}

Analogously, the fine structure related to R2 also exhibits two further absorptions, one peaked at a distance of +56~cm$^{-1}$ from R2 with an
additional shoulder at the low-energy flank and another weaker feature at a distance of
+80.5~cm$^{-1}$ [Fig.~\ref{fig:range1}(b) and (e)]. Again, no zero-magnon feature is discernible, but a direct correspondence to further magnon bands with maxima at about 60~cm$^{-1}$ (7.5~meV) and 80~cm$^{-1}$ (10~meV)\cite{Frontzek2011} can be established by assigning R2 as an exciton-one-magnon sideband at +1\wn of the masked zero-magnon line and the other absorptions to the magnon branches reported by neutron scattering.

In the case of the strong excitation R3 at 19730~cm$^{-1}$ shown in Fig.~\ref{fig:range1}(f) the absorption
spectrum exhibits a broad sideband feature at a distance of +94.5~cm$^{-1}$ from R3, but again no zero-magnon line. Following the same line as above, R3 can be regarded as an exciton-one-magnon sideband corresponding to the small gap of about 1\wn. The assignment of the broad additional sideband is not evident, but it might correspond to an exciton-three-magnon sideband at $m_2+2m_3 \simeq$ 93\wn. Below 18~K the broad sideband becomes distorted and shows an additional shoulder at about +60\wn corresponding to the above mentioned magnon features observed by neutron scattering\cite{Frontzek2011} as in the case of the sideband of R2. This suggests that the sideband is a superposition of two exciton-magnon features.


The excitation R3 exhibits a somewhat different temperature dependence than R1 and R2, because it can be observed up to 45~K (about twice $T_\text{N}$) as a strongly broadened feature in the absorption, while R1 and R2 can not be tracked above $T_\text{N}$. To analyze the temperature dependence of R1 and R2 we fitted the two lines with Gaussians after subtracting the background due to the spin-forbidden excitations by a polynomial fit to the data outside of the exciton range. The
temperature dependence of the fit parameters of R1 and R2 are shown in
Fig.~\ref{fig:fitparam_r1_r2}. Both, the position of the absorption maximum $\omega_0$ [Fig.~\ref{fig:fitparam_r1_r2}(b)] and the intensity [Fig.~\ref{fig:fitparam_r1_r2}(c)]
of the excitations R1 and R2 increase continuously towards lower temperatures reflecting the behavior of the sublattice magnetization in the magnetically ordered state. The line width (full width half maximum) of R1 nearly
triples on rising temperature just before it vanishes, while R2 shows
a less pronounced broadening. Fits of the temperature dependencies of the line width of R1 and R2 by an Arrhenius law
\begin{align}
    \Delta \omega=\Delta \omega_0 + B \exp \left[ \dfrac{-E^*}{k_B T} \right],
\end{align}
yield  the fit parameters $E^*/k_B$ = 31(1)~K, $B$ = 43(3)\wn, $\Delta \omega_0$ = 3.6(1)\wn for R1 and $E^*/k_B$ = 36(2)~K, $B$ = 28(3)\wn, $\Delta \omega_0$ = 11.4(1)\wn for R2.

To analyze R3 we also fitted this range with two Gaussian
lines for R3 and the broad sideband feature [see Fig.~\ref{fig:fitparcucro2}(g)]. The temperature dependence of the fit parameters are shown
in Fig.~\ref{fig:fitparcucro2}. Due to the distortion of the broad sideband below 18~K we only show the fitparameters for R3 in the whole temperature range, while we added the positions of the shoulder and the peak of the sideband to the eigenfrequency below 18~K.
The magnetic ordering temperature shows up in the eigenfrequency $\omega_0$ of R3  [Fig.~\ref{fig:fitparcucro2}(b)] as a kink, below which it increases to lower temperatures, reflecting the increase of an internal magnetic field. A similar behavior is observed for the intensity [Fig.~\ref{fig:fitparcucro2}(c)], but its line width  [Fig.~\ref{fig:fitparcucro2}(a)] seems not to be influenced by the magnetic ordering at $T_\text{N}$. The fit parameters of the broad sideband, which probably corresponds to a superposition of two sidebands, do not exhibit distinct anomalies at $T_\text{N}$.

Previous investigations of the line width of the R3 absorption band
in the related triangular lattice antiferromagnets
LiCrO$_2$ and NaCrO$_2$
\cite{Kojima1993,Elliston1975} have been interpreted in terms of the formation of $Z_2$ vortices,\cite{Kojima1993} i.e., the relaxation time of the magnon sidebands is determined by the density of $Z_2$ vortices for all these compounds. The observed values of the line width in CuCrO$_2$ is comparable to the ones reported for NaCrO$_2$\cite{Elliston1975}. The density of unbound $Z_2$
vortices should be inversely proportional to the correlation length
of two vortices $\xi$ given by\cite{Kawamura2010}
\begin{align}
    \xi=\xi_0 \exp \left[ \dfrac{b}{\left( T/ T_m -1 \right) ^\alpha} \right].
\end{align}
Here $T_m$ corresponds to the melting temperature of the $Z_2$
vortices, i.e., the energy scale to overcome the bound-vortex state, and $\alpha$ is a characteristic exponent.
To check whether this proposal is consistent with our data we use the values for the effective correlation length $\xi_\text{eff}$ determined in neutron scattering measurements by Kajimoto \emph{et al.}\cite{Kajimoto2010}  [Fig.~\ref{fig:fitparcucro2}(a), green circles, right scale]. We find a similar temperature dependence but a stronger increase than the line width of the excitation R3. In addition, we perform a fit of our data using
 \begin{align}
    \Delta \omega=\Delta \omega_m + 1/\xi,
\end{align}
 and
 fixing the values $T_m = 11$~K and $\alpha = 0.37$ to the ones obtained in recent electron spin resonance (ESR) studies of CuCrO$_2$.\cite{Hemmida2009,Hemmida2011} The resulting curve is shown as a solid blue line in Fig.~\ref{fig:fitparcucro2}(a) yielding $\xi_0=2.1(9)\times 10^{-4}$~cm, $b=5.0(6)$, and $\Delta \omega_m$ =14(3)\wn. The data is well described for $T>T_m$ and is consistent with the previous interpretation of the formation of $Z_2$ vortices.  However, an Arrhenius law yields a similarly good fit to the line width of R3 with an activation energy $E^*/k_B$ = 69(1)~K, $B=431(16)\wn$ and $\Delta \omega_0=11.7(4)$\wn (red dashed line). Note that this value is higher than the activation energies of R1 and R2 and higher than the predicted energy cost of the breaking of two $Z_2$ vortices obtained by a Monte Carlo simulation $1.52JS^2= 48$~K.\cite{Okubo2010} A similar analysis of the bands R1 and R2 in terms of  $Z_2$ vortices might be possible between $T_m$ and $T_\text{N}$, but the available temperature range is too narrow to obtain an unambiguous fit.

The putative reason why the excitation R3 has been proposed to be exceptionally susceptible to the existence of $Z_2$ vortices is the fact that R3 and its sideband are observable above the N\'eel temperature and the line width does not seem to be strongly influenced by the onset of long-range magnetic order, but rather reflects the existence of short-range order and spin fluctuations below and above $T_\text{N}$.

\begin{figure}[hb]
\includegraphics[width=0.4\textwidth
]{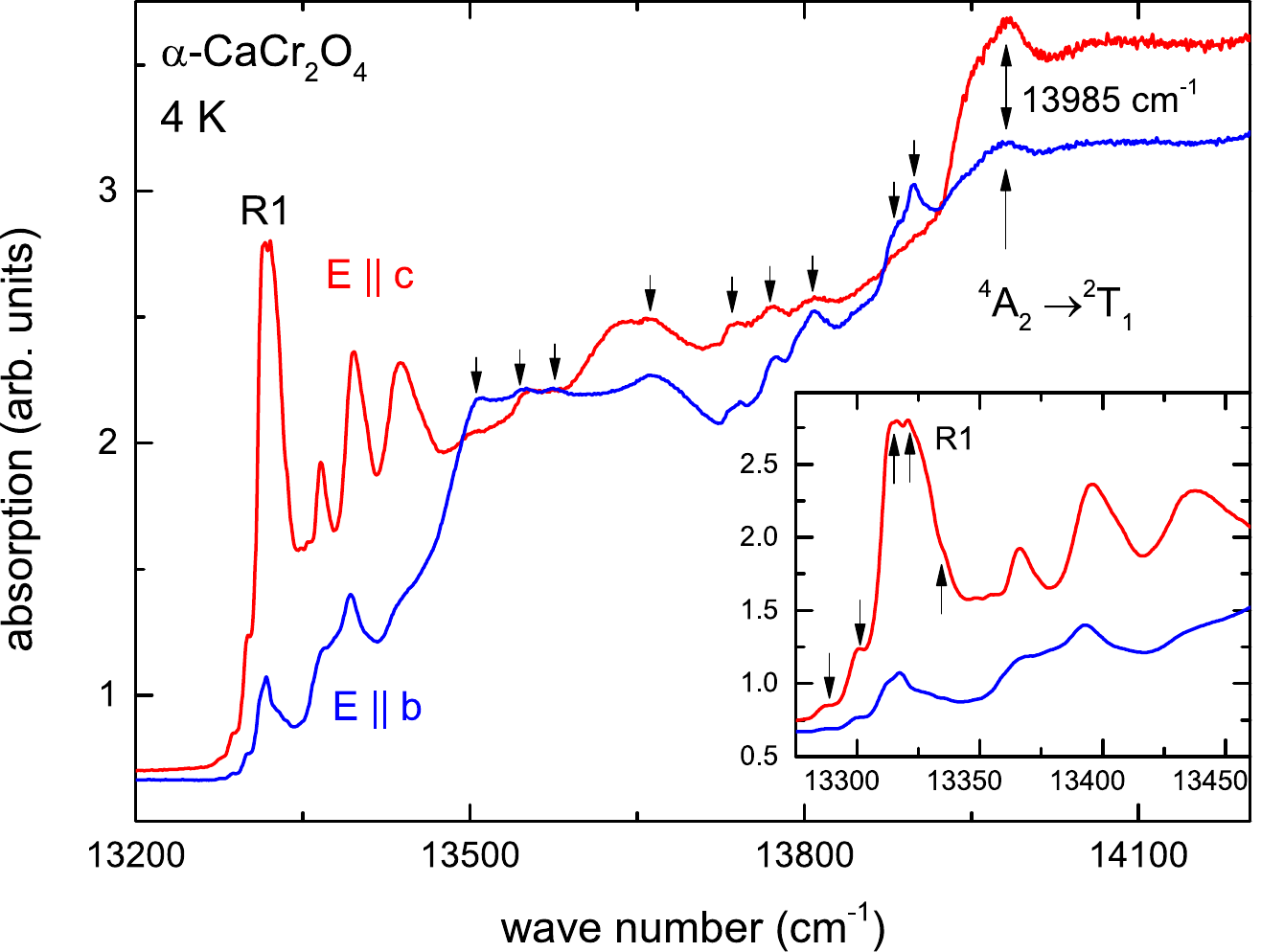}
\caption{\label{fig:cacro_pol_dep} (Color online) Absorption spectra of $\alpha$-CaCr$_2$O$_4$ at 4~K in the frequency region of the $^4A_2 \rightarrow {}^2E$ and $^4A_2 \rightarrow {}^2T_1$
transitions for $E\parallel b$ and $E\parallel c$. }
\end{figure}

\begin{figure*}
\includegraphics[width=0.7\textwidth
]{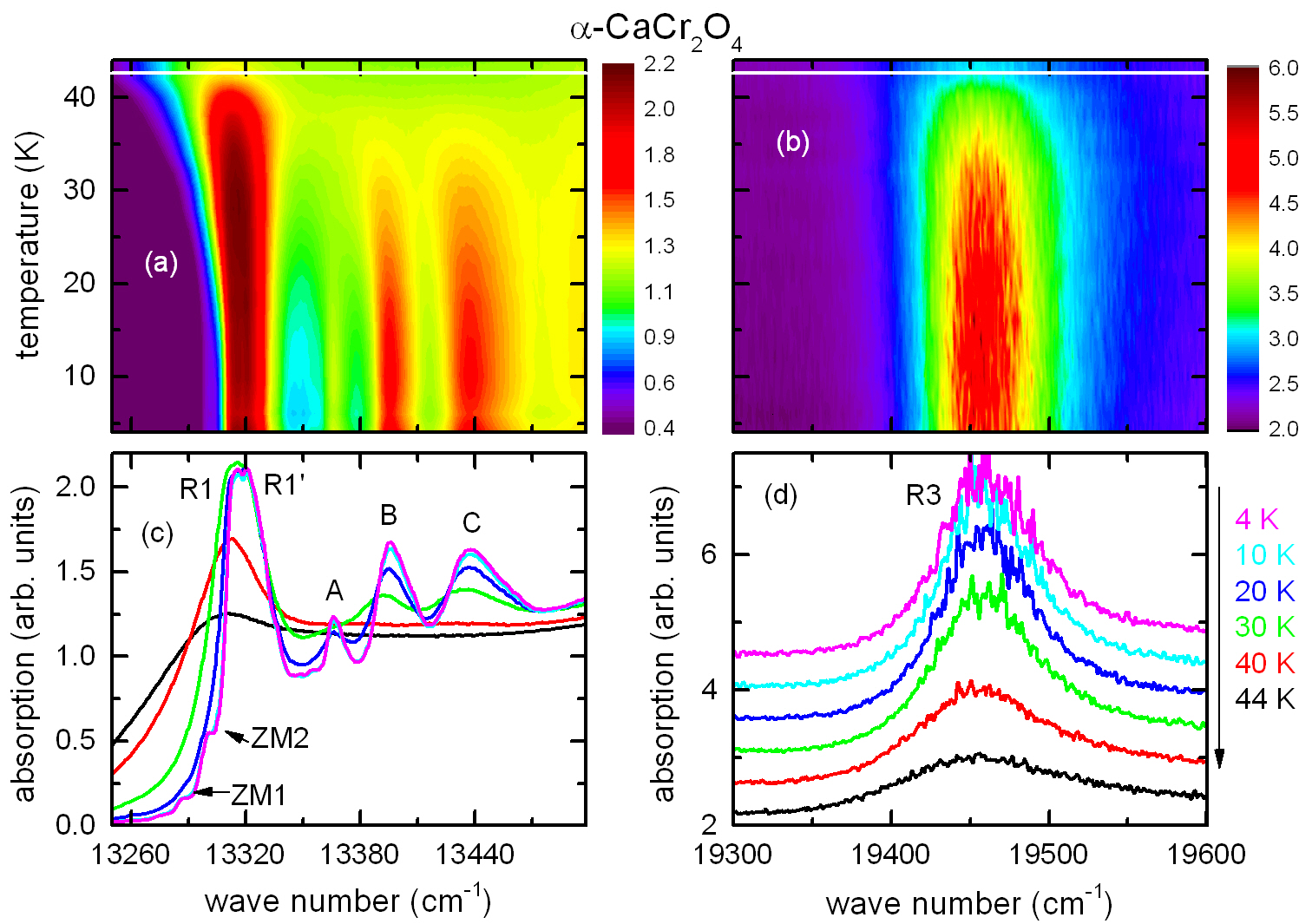}
\caption{\label{fig:trans_contur} (Color online) Color coded contour plot of the absorption as a function of temperature vs.\ wave number in the range
of (a) excitation $^4A_2 \rightarrow {}^2E$ ($E\parallel c$)  and (b) excitation  $^4A_2 \rightarrow {}^2T_2$. $T_\text{N}$ is shown as a white line. Panels (c) and (d) show corresponding absorption spectra for various temperatures in the same frequency range.
The curves are shifted for clarity by a constant value of  0.5 in panel (d).}
\end{figure*}

\begin{figure*}
\includegraphics[width=0.7\textwidth]{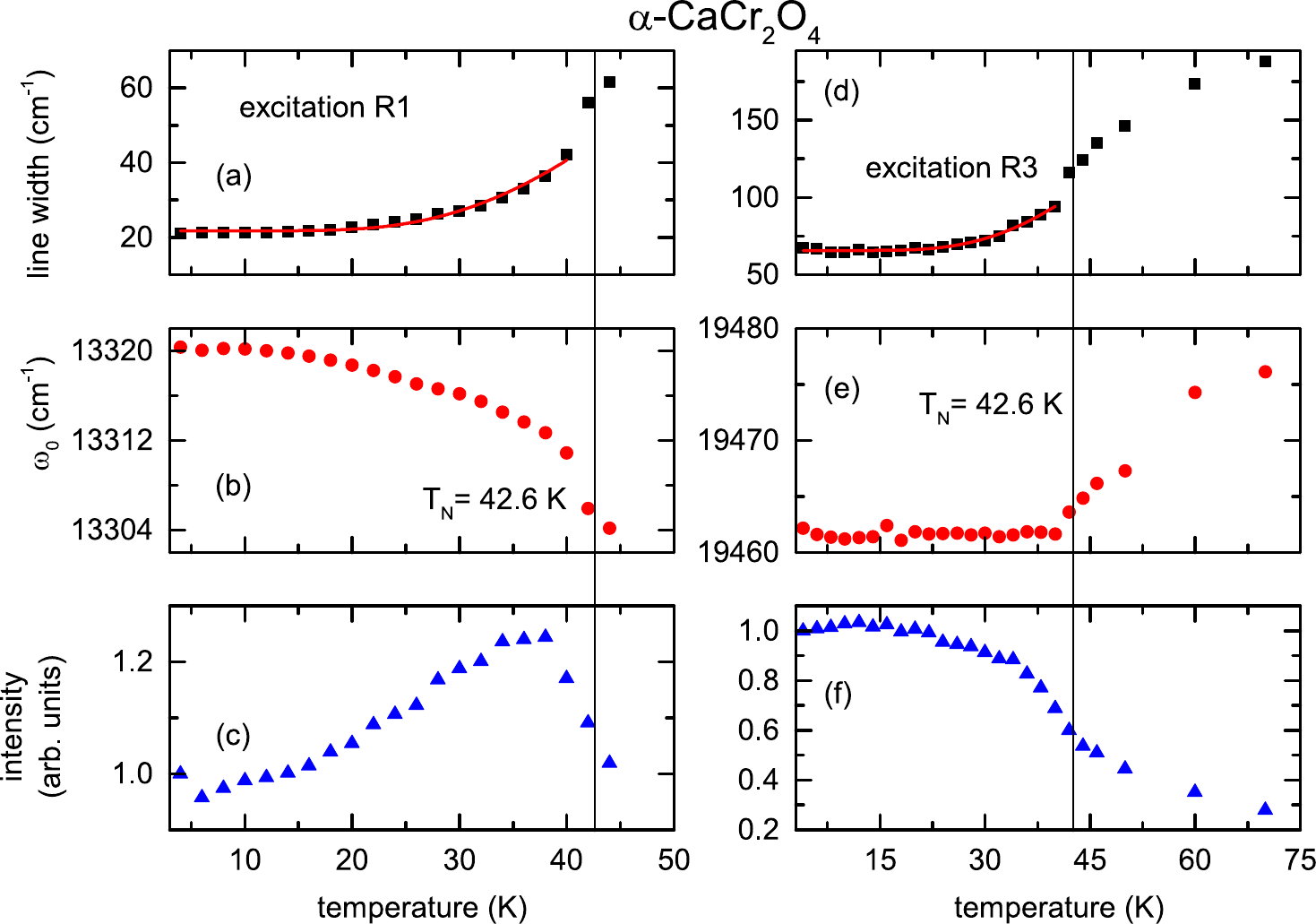}
\caption{\label{fig:fitparcacro} (Color online)  Fit parameters of the absorptions
R1 (left side) and R3 (right side) (a),(d) line width, (b),(e) frequency and
(c),(f) intensity normalized to 4~K, respectively. The solid lines are fits using Eq.~(1) as described in the text.}
\end{figure*}

\subsubsection{$\alpha$-CaCr$_2$O$_4$}

In Fig.~\ref{fig:cacro_pol_dep} the absorption spectra of $\alpha$-CaCr$_2$O$_4$ in the magnetically ordered phase at 4~K are shown in the frequency region of the $^4A_2 \rightarrow {}^2E$ and $^4A_2 \rightarrow {}^2T_1$
transitions for the electric field of the incoming light polarized parallel to the crystallographic $b$ ($E\parallel b$) and $c$ axes ($E\parallel c$). One can clearly observe a pronounced anisotropy, in particular with respect to the fine structure at the onset of the $^4A_2 \rightarrow {}^2E$ transition with the most intense absorption named R1 as in the case of CuCrO$_2$. This fine structure is much more intense for $E\parallel c$, which points toward a corresponding selection rule. The reason that the fine structure is also visible for $E\parallel b$ may be attributed to the fact that the samples reportedly exhibit three twins with one dominating twin (accounting for about 64\% in another sample\cite{Toth2011}).
In comparison to CuCrO$_2$ the spectrum exhibits a more complex fine structure in the plotted frequency regime. In the inset of Fig.~\ref{fig:cacro_pol_dep} arrows indicate weak features around R1 and even a small but resolvable splitting of R1 with maxima at 13316\wn and 13320\wn is visible at the lowest temperatures. In the region where we expect the onset of the $^4A_2 \rightarrow {}^2T_1$ transition with the corresponding absorption R2, there are several weak absorption features appearing for both polarizations (indicated by arrows in Fig.~\ref{fig:cacro_pol_dep}). The most prominent feature at 13985\wn for $E\parallel c$  is tentatively assigned to the onset of the $^4A_2 \rightarrow {} ^2T_1$ transition. Given the complexity of the spectra, we will restrict the discussion
to the temperature dependence of the R1 fine structure (for $E\parallel c$) and the excitation R3 at higher frequencies [see Fig.~\ref{fig:trans_contur}(d)], which does not exhibit a significant polarization dependence.

The corresponding absorption spectra are shown in
Fig.~\ref{fig:trans_contur}(c) and (d) for several temperatures below and
above the N\'eel temperature. The detailed temperature evolution is visualized
in the color-coded plots in Fig.~\ref{fig:trans_contur}(a) and (b). The
emergence of the fine structure in the vicinity of R1 and the appearance of R3 are clearly related to the magnetic ordering below $T_\text{N}$.
In contrast to the case of CuCrO$_2$ we find weak shoulders at the low-energy flank of the most intense mode R1, which we assign to zero-magnon lines ZM1 and ZM2 at 13285\wn and at 13298\wn, respectively. The observation of two zero-magnon-like features and the splitting of R1 are attributed to the fact that there are two inequivalent Cr sites in the structure (see Fig.~\ref{fig:struc}).\cite{Toth2011} The difference in crystal-field strength at the two sites may account for the two different zero-magnon energies. The sideband energies with respect to ZM1 and ZM2 are listed in Table \ref{table1} to compare with the experimentally reported magnon energies. 

The magnetic excitation spectra determined by neutron scattering\cite{Toth2012} revealed van-Hove singularities at 5, 11, 17, and 33 meV in powder samples corresponding to flat regions in the spin-wave dispersions. Moreover, single-crystal studies showed a gap of 3.5~meV (assigned to interplane coupling) and further flat regions at about  8~meV (81\wn). These two energies are in agreement with the distances of the R1 feature from ZM1 and the peak at +80.1\wn (A) from ZM2 and justify their assignment to exciton-one-magnon sidebands. Also for the three low-energy van-Hove singularities we find corresponding sideband energies. Therefore we can assign all the strong absorptions in the vicinity of R1 to exciton-magnon excitations.

The $^4A_2 \rightarrow {}^2T_2$ (R3) band only reveals the absorption R3 at
19450\wn without any sign of a zero-magnon line and, hence, an identification as an exciton-magnon transition is not obvious. The excitation R3 broadens and seems to disappear on approaching $T_\text{N}$
from below, although a broad background contribution  is still visible above
the N\'eel temperature.


Following the same procedure as in the case of CuCrO$_2$, we evaluate the main absorption peaks R1 (neglecting its small splitting at lowest temperatures) and R3 using Gaussian line shapes. The fit parameters are shown in
Fig.~\ref{fig:fitparcacro}. The excitation R1 can be followed up to 44~K.
The eigenfrequency shows the typical increase to lower temperatures and, as expected, the line width increases strongly when approaching the magnetic ordering temperature from below. The intensity increases with temperature and exhibits a broad maximum at around 35~K before it decreases again when magnetic order gets lost. As in the case of CuCrO$_2$ the excitation R3 can be tracked as a very broad feature up to 70~K, but the eigenfrequency is almost constant below $T_\text{N}$, which is not expected for a magnon sideband, and increases for $T>T_\text{N}$. The temperature dependence of the line width of both lines can be described by an Arrhenius law [Eq. (1)] below $T_\text{N}$ and yields $E^*/k_B$ = 152(12)~K, $B$ = 846(250)\wn, $\Delta \omega_0$ = 21.8(3)\wn for R1 and $E^*/k_B$ = 157(11)~K, $B$ = 1455(404)\wn, $\Delta \omega_0$ = 65.8(4)\wn for R3. The values of the activation energy of the two lines are very close, suggesting that the lifetime of both lines is governed by the same mechanism.

\begin{table}[b]
\caption[]{\label{table1}Comparison of the sideband distances with respect to ZM1 and ZM2 and the energies of van-Hove singularities and flat regions of the spin-wave dispersion reported in Ref.~[\onlinecite{Toth2012}]. The energies are given in meV. Discarded assignments are given in parentheses.}
\quad

\begin{tabular}{l|c|c|c}
                & ~~ZM1~~ & ~~ZM2~~  &~Ref.~[\onlinecite{Toth2012}]~   \\
  \hline
  R1  &      3.7     &     (1.8)       &   3.5       \\
  R1$^\prime$~  &      4.3     &     (2.4)       &   5       \\
  A  &      (9.9)    &     8.1       &    8      \\
  B  &      (13.6)    &     11.8      &   11      \\
  C  &     (18.8)    &     17.0        &      17      \\

\end{tabular}
\end{table}

\begin{figure*}[t]
\includegraphics[width=0.8\textwidth]{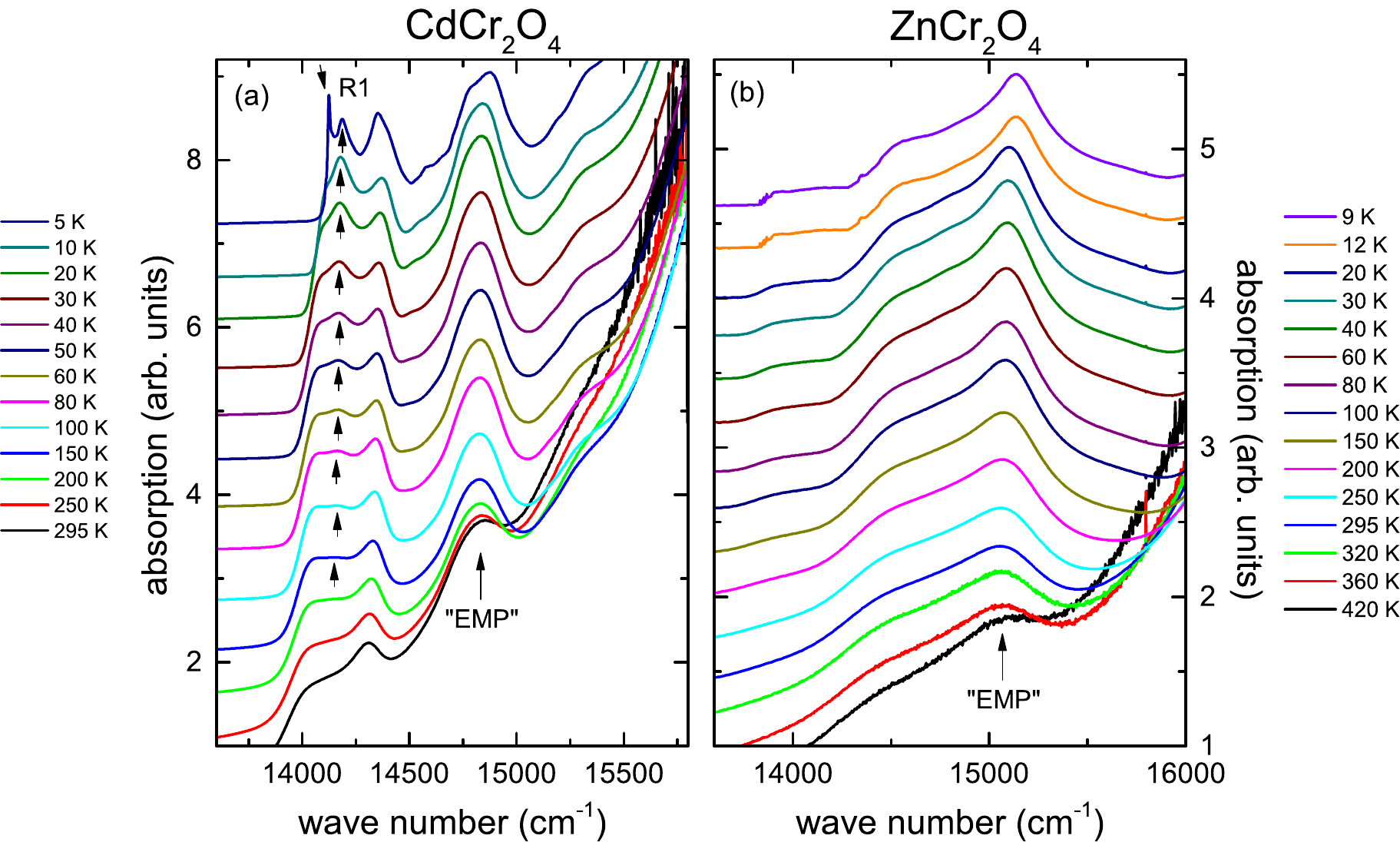}
\caption{\label{fig:detail_cco_zco} (Color online) Absorption spectra of CdCr$_2$O$_4$ (a) and ZnCr$_2$O$_4$ (b) for various temperatures in the energy region of the excitations within the $t_{2g}$ levels. The spectra are shifted for clarity by 0.55 (CdCr$_2$O$_4$) and 0.29 (ZnCr$_2$O$_4$).}
\end{figure*}

\subsubsection{CdCr$_2$O$_4$ and ZnCr$_2$O$_4$}

The absorption spectra in the region of the spin-forbidden $^4A_2 \rightarrow {}^2E$ and $^4A_2 \rightarrow {}^2T_1$ crystal-field excitations of CdCr$_2$O$_4$ and ZnCr$_2$O$_4$ are shown for several temperatures in Fig.~\ref{fig:detail_cco_zco}(a) and (b), respectively.

\begin{figure}[h]
\includegraphics[width=0.49\textwidth]{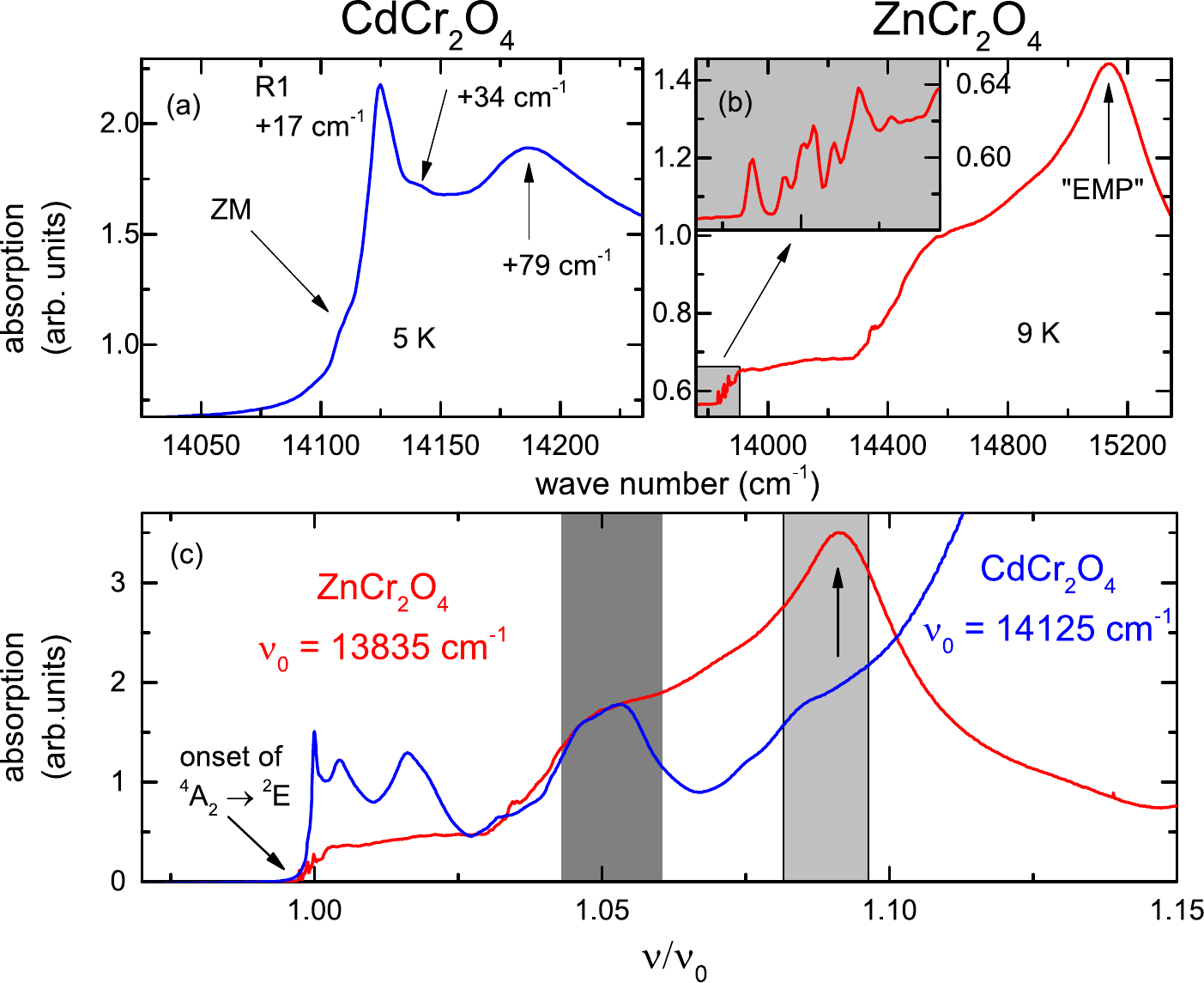}
\caption{\label{fig:zoom_cco_zco} (Color online) Zoom into the lower energy part of the absorption spectrum of CdCr$_2$O$_4$ (a) and ZnCr$_2$O$_4$ (b).  For the latter the grey part is shown in an inset in an extended view. (c) shows a scaled version of the spectra. For details of the scaling see text.}
\end{figure}

 The temperature evolution of the spectra in ZnCr$_2$O$_4$ is rather continuous and the broad absorption peaks sharpen with decreasing temperature as expected in case of phonon-assisted transitions. In addition, the smooth onsets of the transition bands sharpen up and below the magneto-structural transition the onset is edge-like and exhibits a very weak fine structure. The lowest-temperature spectrum is shown in an enlarged scale in Fig.~\ref{fig:zoom_cco_zco}(b). This group of narrow excitonic lines (further zoomed in the gray inset) has previously been observed by Szymczak \emph{et al}.\ and attributed to the Davydov
splitting of the two lowest single-ion levels of the $^2E$ state.\cite{Szymczak80} Interestingly, very similar spectra have also been obtained and analyzed for exchange-coupled Cr$^{3+}$ pairs doped into non-magnetic ZnGa$_2$O$_4$.\cite{Gorkom1973} These authors identified the observed fine structure in terms of exciton-magnon processes and vibronic sidebands. Remarkably, the obtained values for the exchange coupling constants in dilute systems are in very good agreement with the ones obtained for ZnCr$_2$O$_4$ itself.\cite{kant09a, kant10a} Above the magnetic ordering temperature this fine structure is smeared out.

The rather broad absorption band in ZnCr$_2$O$_4$ at
15150~cm$^{-1}$ associated with $^4A_{2}\rightarrow {}^2T_{1}$ has been suggested to
stem from an exciton-magnon-phonon (EMP) transition,\cite{Szymczak80} which has recently been reported to be suppressed in fields above 400~T.\cite{Miyata2011} We traced its temperature evolution to temperatures above the Curie-Weiss temperature as shown in Fig.~\ref{fig:detail_cco_zco}(b) and still observe a broad maximum at the highest measured temperatures. This suggests that this excitation might not be governed by spin correlations, but rather corresponds to a more conventional CF excitation. Note that a very recent high-magnetic field study of this frequency region in CdCr$_2$O$_4$ reports the suppression of both the fine structure around R1 and the absorption assigned to an EMP transition in analogy to ZnCr$_2$O$_4$.\cite{miyata2013}

Concerning the excitation R3 we did not observe any narrow absorption features for CdCr$_2$O$_4$ in this region and for ZnCr$_2$O$_4$ only a rather broad feature (not shown) has been observed, which might indicate the spin-forbidden  $^4A_{2}\rightarrow {}^2T_{2}$ transition.

In the case of CdCr$_2$O$_4$ one can clearly see the emergence of the absorption peak R1 at 14125\wn related to the $^4A_2 \rightarrow {}^2E$ transition. In an enlarged scale [see Fig.~\ref{fig:zoom_cco_zco}(a)] shoulders are visible in the spectrum at a distance $\pm$17(2)\wn from R1. In line with the discussion of CuCrO$_2$ and $\alpha$-CaCr$_2$O$_4$ the shoulder at the low-energy flank is assigned to a zero-magnon line ZM and R1 to an exciton-one-magnon sideband. The magnetic excitation spectrum of CdCr$_2$O$_4$ has been studied by neutron scattering and spin wave excitations with energies $m_a=0.65$~meV (5.2\wn), $m_b = 2.3$~meV (19\wn) and $m_c = 4.7$~meV (38\wn) were reported at the wave vector (1,0.915,0) associated with the zone center of the incommensurate spin structure.\cite{Chung2005} In addition, in a high-field ESR study up to ten magnetic excitations have been observed and attributed to higher-harmonics of helical spin-resonance modes.\cite{Kimura06} A recent theoretical analysis confirmed this idea and described the magnetic excitation spectrum as due to helimagnons.\cite{Choi2013, Choi2013a} We identify R1 as an exciton-one-magnon absorption involving the magnon energy $m_b$, which reportedly is only slightly dispersive.\cite{Chung2005} The weak shoulder at a distance of +34\wn could then correspond to an exciton-one-magnon sideband involving $m_c$.  Both the weak shoulders and mode R1 can not be detected anymore above the magnetic ordering temperature.

A further absorption peak is visible at a distance of +79\wn from ZM and remains visible above the N\'eel temperature and persists up to about 150~K, where the absorption spectrum seems to become plateau-like [see arrows in Fig.~\ref{fig:detail_cco_zco}(a)].
This behavior is in contrast to the other absorption features at higher frequencies, which remain visible up to room temperature and, therefore, are likely not directly related to spin correlations. We speculate that the peak at +79\wn might therefore arise due to short-range spin correlations. A possible assignment of this peak could be an exciton-three-magnon peak involving $2m_b +m_c$.


A direct comparison of the two low-temperature absorption spectra of CdCr$_2$O$_4$ and ZnCr$_2$O$_4$ is presented in Fig.~\ref{fig:zoom_cco_zco}(c), where the spectra are plotted vs.\ wave number scaled to the respective onset $\nu_0$ of the exciton lines. In case of CdCr$_2$O$_4$ $\nu_0$ was taken as the well-defined frequency of the excitation R1 and in case of ZnCr$_2$O$_4$ the first peak of the fine structure at 13835~cm$^{-1}$ was used. The absorption values are scaled in a way that the curves coincide at the maximum of the absorption peak visible at $\nu/\nu_0$ = 1.05 for CdCr$_2$O$_4$. In this scaling, even the low-energy flanks of the two curves coincide and the peak maximum in CdCr$_2$O$_4$ corresponds to a clear shoulder in ZnCr$_2$O$_4$. Moreover, the EMP feature in ZnCr$_2$O$_4$ seems to correspond to a shoulder in the spectrum of CdCr$_2$O$_4$ [see shaded areas in Fig.~\ref{fig:zoom_cco_zco}(c)]. This shoulder is directly followed by the first spin-allowed crystal-field excitation in CdCr$_2$O$_4$. We propose that the occurrence of similar features in both compounds and their persistence to the highest studied temperatures (exceeding the respective Curie-Weiss transitions) again favors a structural origin of these excitations. Therefore, the reported suppression of the EMP peak above 400~T in ZnCr$_2$O$_4$ might be interpreted as a magnetic-field induced structural change.\cite{Miyata2011}
 This scaling further suggests that the $^4A_2 \rightarrow {}^2E$ and $^4A_2 \rightarrow {}^2T_1$ regions are behaving differently in terms of relative intensities for the two compounds, i.e., the $^4A_2 \rightarrow {}^2E$ and $^4A_2 \rightarrow {}^2T_1$ regions appear to be of comparable intensity in CdCr$_2$O$_4$, while the $^4A_2 \rightarrow {}^2T_1$ region seems to have much higher intensity in  ZnCr$_2$O$_4$. This might be related to the different structural distortions in the two systems in the magnetically ordered state. There is still a debate with respect to the exact symmetry of the two systems,\cite{Chern06,Ji09,Kant09} but
a clear difference is that in  ZnCr\tsub2O\tsub4 the lattice contracts and
in CdCr\tsub2O\tsub4 it elongates along the $c$-axis. This will certainly influence the transition probabilities and the intensities of the CF transitions. 
The difference in the exciton-magnon features for the two spinel systems is in our opinion due to the fact that the ground states of CdCr$_2$O$_4$ and ZnCr$_2$O$_4$ differ strongly. In CdCr$_2$O$_4$ a helical magnetic structure seems to be favored, while the complex magnetic structure of ZnCr$_2$O$_4$ has not yet been solved and might be dominated by weakly coupled molecule-like units.

\section{Summary}

We investigated the Cr crystal-field excitations in the triangular-lattice
antiferromagnets CuCrO$_2$ and \mbox{$\alpha$-CaCr$_2$O$_4$} and in the spinels
CdCr$_2$O$_4$ and ZnCr$_2$O$_4$ in order to search for exciton-magnon transitions as optical probes of the magnetic excitation spectra, the existence of short-range spin correlations, and anomalies in the lifetime of magnons due to frustration effects.
In particular, optical transmission experiments can be performed on small samples and detailed temperature dependencies can be obtained. Moreover, probing exciton-magnon sidebands in pulsed high-magnetic fields can provide information on the magnetic excitation spectrum, which is inaccessible by neutron scattering (see e.g.\ Ref.~[\onlinecite{Miyata2011},~ \onlinecite{miyata2013}]).

In CuCrO$_2$ no zero-magnon lines could be observed, but three sets of fine structures have been detected to emerge
with lowering temperature to the magnetically ordered state. The measured absorption peaks
were assigned to exciton-magnon transitions by comparison with magnon energies
found by antiferromagnetic resonance and neutron scattering experiments assuming that the zero-magnon line is almost coinciding with and therefore masked by the first intense exciton-one-magnon sideband. The temperature dependence of the line width of the most intense exciton-one-magnon sidebands R1-R3 can be described by an Arrhenius law, but yield different activation energies. The sidebands R1 and R2 related to $^4A_2 \rightarrow {}^2E$ and $^4A_2 \rightarrow {}^2T_1$ can only be resolved below the N\'eel temperature, but the sideband R3 associated with the $^4A_2 \rightarrow {}^2T_2$ crystal-field transition can be tracked even above $T_\text{N}$ and its line width does not exhibit an anomaly at the N\'eel temperature. This suggests that the line width of R3 can be regarded as a probe of short-range spin correlations which already exist above $T_\text{N}$. Due to this fact the temperature dependence of the line width of the excitation R3 in related systems has previously been described by assuming that the exciton-magnon lifetime is determined by the existence of $Z_2$ vortices. We find that a similar analysis holds also for CuCrO$_2$.

In the other triangular antiferromagnet $\alpha$-CaCr$_2$O$_4$ we observed an even richer fine structure related to the $^4A_2 \rightarrow {}^2E$ and $^4A_2 \rightarrow {}^2T_1$ crystal-field excitations which depend on the polarization of the light with respect to the crystallographic axis in the investigated plane. Probably due to the twinning of the crystal a strict selection rule could not be observed, but the fine structure at the onset of the $^4A_2 \rightarrow {}^2E$ transition including R1 was much stronger for $E\parallel c$. In contrast to CuCrO$_2$ weak zero-magnon lines were found below R1 and the sidebands could be directly related to spin-wave energies reported by neutron scattering. Again the magnon sidebands around R1 disappear on approaching the N\'eel temperature from below, but the excitation R3 remains  visible also above $T_\text{N}$ as in the case of CuCrO$_2$. The line widths of both lines follow an Arrhenius law with a similar activation energy, but both, intensity of R1 and eigenfrequency of R3, show an unexpected behavior and indicate that the exciton-magnon properties are more complex than in CuCrO$_2$. Although we can not exclude the existence of $Z_2$ vortices in this compound, a description of the line width of R3 across the N\'eel temperature in terms of a $Z_2$ vortex scenario is not possible.

Even though these two triangular systems exhibit an almost ideal 120$^\circ$ helical structure the exciton-magnon features differ strongly. The reason for this difference is probably the orthorhombic distortion of $\alpha$-CaCr$_2$O$_4$ and different Cr sites. This results in a more complex spin-wave spectrum and inequivalent exchange interactions between the Cr ions driving the system further away from the isotropic nearest neighbor Heisenberg case.

In CdCr$_2$O$_4$ a zero-magnon line and exciton-magnon sidebands related to the $^4A_2 \rightarrow {}^2E$ transition have been
observed at lowest temperatures in agreement with the reported spin-wave dispersions by neutron studies. These features disappear concomitantly with long-range magnetic order, but another sideband at +79\wn remains visible up to about 150~K.

In contrast, in the spinel ZnCr$_2$O$_4$ only a weak fine structure has been
observed in agreement with a previous report. The temperature dependence of
an absorption (EMP) reportedly associated with the magnetic structure and
suppressed in very high-magnetic fields suggests a primarily structural
origin of this excitation.

The helical magnetic structure in CdCr$_2$O$_4$ with its well defined magnon branches and exciton-magnon features is in contrast to the still unresolved  magnetic ground state of ZnCr$_2$O$_4$ and the proposed existence of weakly coupled multi-spin entities. Such molecular units might produce the previously observed Davydov splitting\cite{Szymczak80} instead of strong exciton-magnon sidebands.

\begin{acknowledgments}
We want to thank M.V. Eremin, N. Perkins, H.-A. Krug von Nidda, M. Hemmida, and O. Tchernyshyov for fruitful discussions.
We acknowledge partial support by the Deutsche
Forschungsgemeinschaft via TRR 80 (Augsburg-Munich)
and project DE 1762/2-1.
\end{acknowledgments}

\end{document}